\documentclass[a4paper, 12pt]{article}

\usepackage{graphicx}  
\usepackage{bm}        
\usepackage{amssymb}   
\usepackage{amsmath}

\newcommand{\intervalleff}[2]{\mathopen{[}#1\,;#2\mathclose{]}}

\begin{document}

\title{Infrared light emission from atomic point contacts}

\author{T. Malinowski, H.R. Klein, M. Iazykov, Ph. Dumas\\
Aix Marseille Universite, CNRS, CINaM UMR 7325\\ 13288, Marseille, 
France\\
klein@cinam.uiv-mrs.fr
}




\maketitle

\begin{abstract}
Gold atomic point contacts are prototype systems to evidence ballistic electron 
transport. The typical dimension of the nanojunction being smaller than the 
electron-phonon interaction length, even at room temperature, electrons transfer 
their excess energy to the lattice only far from the contact. At the contact 
however, favored by huge current densities, electron-electron interactions 
result in a nano hot electron gas acting as a source of photons. Using a home 
built Mechanically Controlled Break Junction, it is reported here, for the first 
time, that this hot electron gas also radiates in the infrared range (0.2eV to 
1.2eV). Moreover, in agreement with the pioneering work of 
Tomchuk \cite{Tomchuk1966},we show that this radiation is compatible with a 
blackbody like spectrum emitted from an electron gas at temperatures of several 
thousands of Kelvin given by $(kB.Te)^2 = \alpha. I.V$ where $\alpha$, $I$ and 
$V$ are respectively a fitting parameter, the current flowing and the applied 
bias.
\end{abstract}

\section{Introduction}
Understanding and managing the interplay between electrons and photons 
around the Fermi level is of paramount importance for both fundamental and 
applied solid states physics. The recent intense research works in the field of 
nanoantennas \cite{Novotny2011} or regarding light-emitting diode droop see 
\cite{Iveland2013}, and references therein. illustrate this importance. As the 
sizes of the active regions shrink to the nanometre scale, as local current 
densities increase, new processes, previously not favoured, are put forward, 
intentionally or not.

The tip of a scanning tunneling microscope has been used to 
inject electrons and to promote local light emission from semiconductor quantum 
structures \cite{Renaud1991,Ushioda2000,Dumas2000} or from single molecules 
\cite{Qiu2003}. These STM light emission (STM-LE) works followed the pioneering 
work of J.Gimzewski's group on metals \cite{Gimzewski1988,Gimzewski1989}. For 
metals, in the $10^{-4}\ G_0$ conductance range, the well-accepted dominant 
one-electron mechanism is the following \cite{Berndt1991}: an inelastic 
tunneling electron excite collective electron modes of the tip-gap-surface 
nanocavity. These modes depend on the geometry of the cavity at the nanoscale 
and on the dielectric properties of the metals \cite{Rendell1981}. These 
electromagnetic modes relax their energy mainly to the phonons but also through 
photon emission. The two key features of the emitted spectra are i) that they 
exhibit plasmonic resonances typical of the cavity and metals-dependent, and 
ii) that the high energy part of the spectra is limited by the energy carried by 
a tunneling electron ($h\nu \leqslant eV$) \cite{Coombs1988}. Although it is 
obviously not possible to know the tip shape at this scale, and thus the 
electromagnetic modes due to the tip-gap-surface nanocavity, it has been shown 
that rationalizing spectra acquired with the same tip on different areas or at 
different bias conditions could provide useful physical information respectively 
on the material below the tip \cite{Downes2002a,Qiu2003} or on the carrier 
density \cite{Schneider2011}.
 
Indeed photons with energies exceeding the so-called quantum cutoff "limit" of 
$h\nu = eV$ have also been observed in STM-LE regime 
\cite{Pechou1998,Hoffmann2003}. Such photon energies are still observed at 
higher conductances, above $G_0$ in the Atomic Point Contact Light Emission 
(APC-LE) regime \cite{Downes2002,Schull2009}. The emission of these photons 
evidences the importance of multi-carrier excitation processes.

During the last decades, atomic sized metallic conductors have been extensively 
studied \cite{agrait2003}, a prototype system being the well-known stretched 
gold nanowire. Under electrical polarization, prior being broken, the 
conductance of such a wire exhibits characteristic Landauer plateaus at integer 
multiples of $G_0=2e^2/h$ \cite{Yanson1998, Klein2008}. Along these plateaus 
the conductance remains constant despite the length increase of the metallic 
nanowire. Indeed, while the length of the nanoconstriction is much smaller than 
the electron-phonon interaction length $L_{e-ph}$ \cite{Ashcroft1976}, no 
extra-resistance is added to the contact resistance. Moreover, considering only 
electron-phonon scattering, the electron injected from one contact to the other 
will preserve his energy and momentum over ballistic distances of the order of 
$L_{e-ph}$. The order of magnitude of $L_{e-ph}$ is given by: 

\begin{equation}
L_{e-ph} = \frac{v_F}{\omega_D \gamma} 
\end{equation}

with $v_F$ the Fermi velocity, $\omega_D$ Debye frequency and $\gamma$, the 
electron-phonon coupling factor ($\gamma<1$) \cite{Ashcroft1976}.
However, these gold nanoconstrictions are the siege of huge current densities 
($|\vec{j}|\simeq10^{15}$ A.m$^{-2}$) and electron-electron interactions 
play a significant role in redistributing the energy of the electrons 
\cite{Pierre2000}.

As mentioned above, APC-LE is also observed. The spectra show the presence of 
photons with energy $h\nu$ above the polarisation energy eV of electrons 
\cite{Downes2002,Schull2009,Buret2015}. The emission of these photons evidences 
the importance of multi-carrier excitation processes resulting in a hot carriers 
energy distribution spreading over more than $eV$.
	
From their first observations, A. Downes \cite{Downes2002} have put 
forward the radiative emission from of a hot electron gas. Consistently with 
previous works on systems with analogue physics by 
Fedorovitch (see \cite{Fedorovich2000} and references therein), electron 
temperatures of the order of $2000$ K were extrapolated fitting the corrected 
emission spectra by a black-body behaviour. Applying a 1 volt bias, at a 
conductance of $1\ G_0$, photons with energy above 2.5 eV were detected. 
Although most spectra were featureless in the visible range, modulation or 
intense peaks, evoking electromagnetic resonances of the tip-gap-surface 
nanocavity were sometimes observed. These experiments were performed with an STM 
in ultra high vacuum (UHV), at $300$ K. In similar conditions, but at 4 K, 
G. Schull \cite{Schull2009} also reported light emission above the quantum 
cutoff. However, their results are different from two important points of view: 
i) the spectra exhibit resonance features similar to what is commonly observed 
in STM-LE and ii) no photon of energy above twice eV is observed. The high 
energy part of the spectra is also attributed to hot electrons, hotter than eV, 
excited through an Auger-like two charge carriers cascade mechanism. A 
mechanism consistent with photon energies between eV and 2eV.
	
Recently, M. Buret \cite{Buret2015} also reported black-body like emission 
from electroformed gold junction at conductance values of the order of $G_0$. As 
in \cite{Downes2002}, they also observed photons with energy above $2eV_{bias}$ 
consistent with a black-body like radiation of an hot electron gas. 
Electronic temperatures, $T_e$, above 1500 K, i.e. well above the gold melting 
point ($T_m = 1338$ K), are indirectly measured. They propose a mechanism 
involving gold interband reabsorption by low-lying d-band electrons to explain 
the apparent experimental discrepancy with \cite{Schull2009}.
	
Using a home-build Mechanical Controlled Break Junction (MCBJ) \cite{Alwan2013}, 
we have been revisiting APC-LE both in the visible range and, for the first 
time, in the near infra-red (IR) range of the spectrum. This article focuses on 
the IR range. We report intense IR emission, conterbalancing the known 
relatively poor sensitivity of IR detectors. We also report basic spectroscopic 
data, supporting a blackbody-like emission from hot electron gas.
	
One of the reasons for focusing on the emission in the IR range is that we do 
not expect electromagnetic plasmonic resonances comparable to what is observed 
in the visible range. In the classical theory \cite{Rendell1981} we would expect 
a diverging redshift of these resonances as the distance between electrodes is 
reduced from the STM regime down to the contact regime. Noteworthy, to our 
knowledge, this redshift was never observed. Indeed, recent quantum approaches 
have theoretically predicted \cite{Romero2006,Savage2012} and experimentally 
demonstrated \cite{Savage2012} a non monotonous behaviour, limiting the 
wavelength of the resonances below 1 micron.

\section{Experimental set-up}
	
\begin{figure}[!h]	

\centerline{\includegraphics[width=70mm,height=90mm]{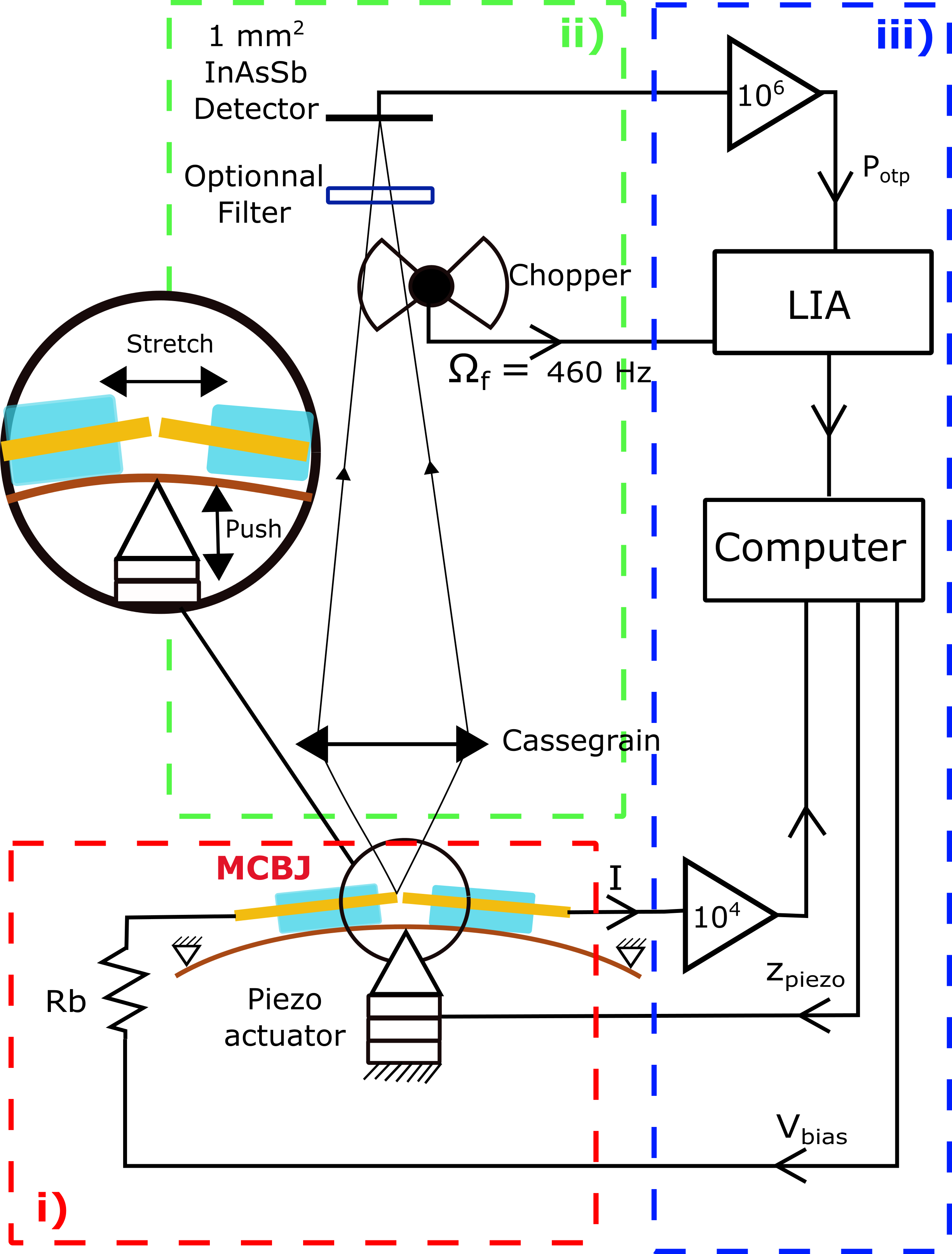}}
\caption{ Experimental set-up. i) MCBJ with a piezo actuator. Photons 
are collected using a reflective Cassegrain objective (15x, N.A.=0.5), 
optionnally filtered and chopped at $460\ Hz$ and detected using a cooled InAsSb 
IR detector ii). The magnified view shows the MCBJ principle (push-to-stretch 
movement). The control and acquisition electronics is limited to current-voltage 
convertors, a lock-in amplifier (LIA) and a computer iii).}
\label{fig : Sch}
\end{figure}

\begin{figure}[!h]	
\begin{center}
\includegraphics[width=85mm,height=35mm]{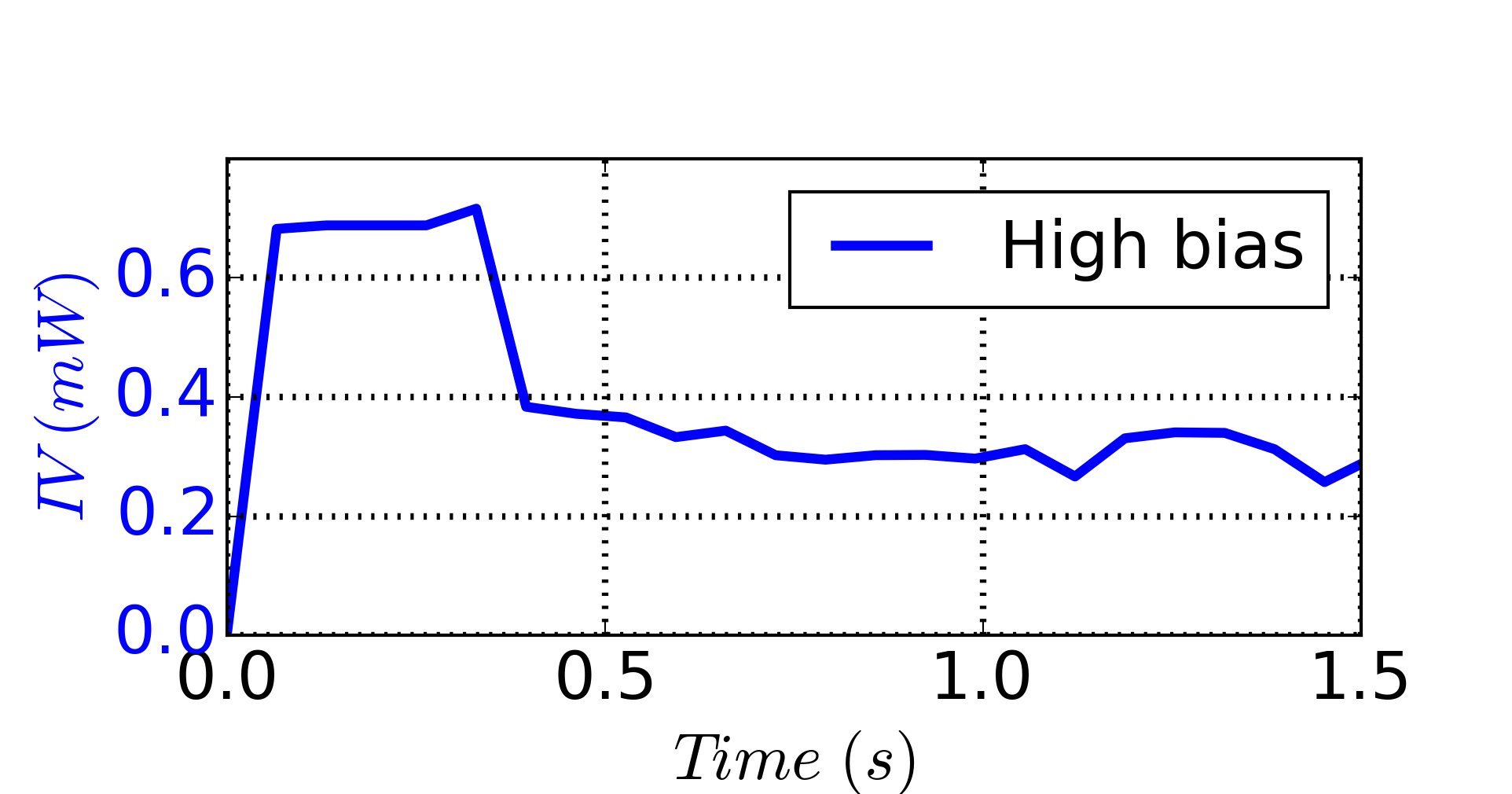}\\
\includegraphics[width=85mm,height=35mm]{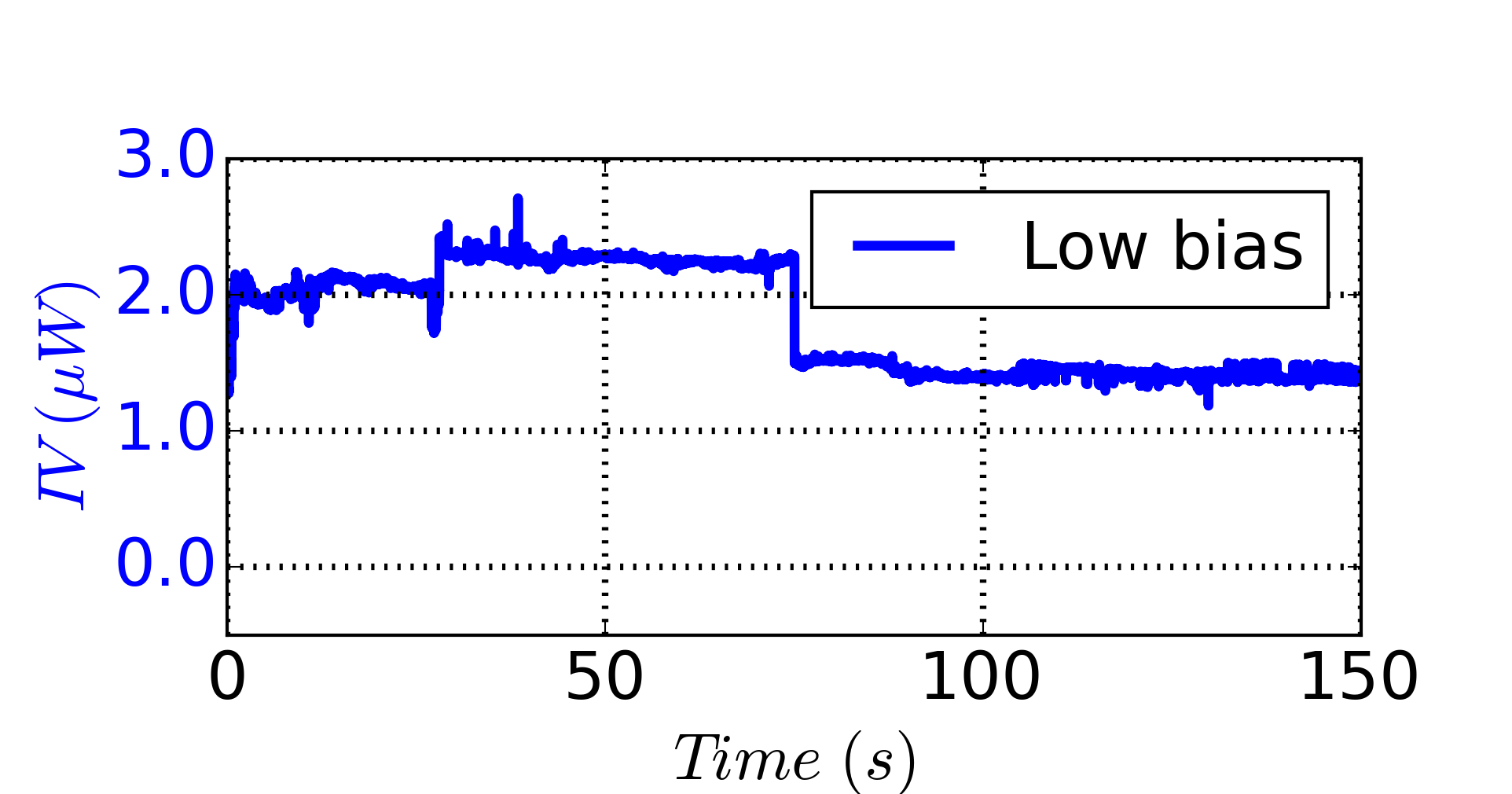}\\
\end{center}
\caption{Temporal evolution of $IV$, the electrical power, without feedback. 
\textbf{Top:} high bias regime ($V_{bias}=1.5$ V). \textbf{Bottom :} low bias 
regime ($V_{bias}=0.139$ V).}
\label{fig : IR4}
\end{figure}

For these experiments, the setup consists of i) a MCBJ ii) the light collection 
and detection components and iii) the acquisition and control electronics and 
informatics (see fig.\ref{fig : Sch}).
The MCBJ was first introduced by Ruitenbeek et al \cite{Ruitenbeek1996}. For 
these studies, our is operated in air and at room temperature. The mechanical 
part is similar to the one we described previously \cite{Alwan2013}, although 
the sample preparation technique has been since improved. The separation of the 
electrodes is controlled by a micrometer step motor stacked-up with a 
piezoelectric actuator (sensibility : $216\ nm.V^{-1}$). Motor and piezo are 
driven through an input/output (IO) board by a computer interface (written with 
Labview)  that is also used for acquiring data and feedback (see below). Taking 
into account a typical push:stretch ratio of 20:1 and the resolution of our 
16-bit DAC, one digit corresponds to less than $3\ pm$ which is quite enough for 
this work.\\
The junction, in series with a $1\ k\Omega$ ballast resistor is biased from the 
IO board. The conductance is derived from the measured intensity that flows 
through the junction using a current/voltage converter (DLPCA-200, FEMTO) with 
a $10^4$ A/V transconductance gain.
At low bias ($V_{bias} \simeq 130\ mV$ ), in air and at room temperature, atomic 
contacts often remain stable for tens of seconds \cite{Alwan2013}. 
Figure\ref{fig : IR4} illustrates the long term stability of junctions biased 
at low voltage. As the bias is increased up to values stimulating light 
emission, the lifetime of monoatomic contacts decreases drastically below $100\ 
ms$.
	
To collect infra-red photons, we are using a Cassegrain microscope objective 
(x15 ; $NA=0.5$). The optical beam is mechanically chopped at $460\ Hz$, 
transmitted through a semiconductor filter and measured by a cooled InAsSb 
detector (P11120-201, Hamamatsu) sensitive from $0.2\ eV$ to $1.2\ eV$, 
using a lock-in amplifier (HF2LI, Zurich Instruments) . We use silicon 
and germanium wafers of respective gap $1.12\ eV$ and $0.68\ eV$ as lowpass 
filters to gather first-order spectroscopic data.

To measure an optical IR signal we operate with an input electrical power in the 
mW range. More precisely, we apply a bias in the volt range and drive the MCBJ 
at conductance of a few $G_0$. The MCBJ device is mechanically and thermally 
stable at the macroscopic scale. Moreover, taking advantage of thermal diffusion 
and electromigration, at room temperature, the nanojunction self-organises at 
atomic level and naturally explores the more stable configurations around the 
average chosen conductance value. We thus only need a loose feedback using the 
piezo actuator to maintain the conductance between $\intervalleff{\ 0.5\ G_0\ 
}{\ 20\ G_0\ }$.

\section{Results}

While we deliberately drive our MCBJ in a loose feedback mode, its conductance 
naturally varies during the experiments. We have plotted (figure \ref{fig : 
IR4}) the fluctuation of electrical power $IV$, which is a relevant parameter 
\cite{Fedorovich2000}, as a function of time. It shows that the applied bias and 
the stability are negatively correlated. A compromised to measure an optical 
signal thus has to be found.

\begin{figure}[!h]	
\centerline{\includegraphics[width=90mm,height=50mm]{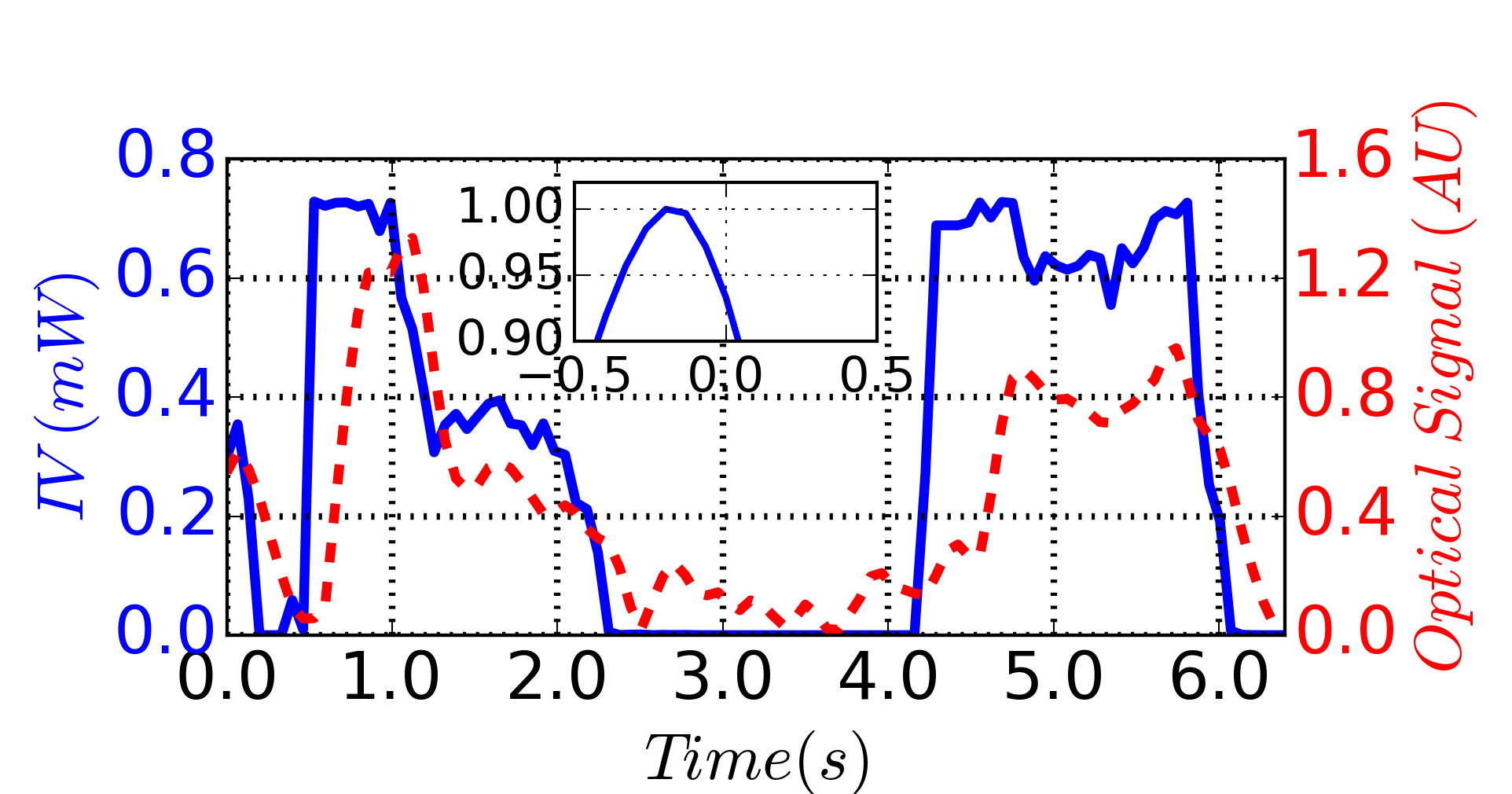}}
\caption{ Temporal evolutions of the electrical power $IV$ (blue line) 
and optical signal (red dashed line). The delay is due to the chosen LIA time 
constant. It can be quantified by the cross-correlation (see inset).}
\label{fig : IR3}
\end{figure}

Figure.\ref{fig : IR3} plots together the temporal evolution of the measured 
electrical power and optical signal. The optical response, measured without 
optical lowpass filters, appears to be strongly correlated with the electrical 
power injected in the junction and slightly time-delayed. Cross-correlation 
(figure \ref{fig : IR3}, inset) of both signals allows to quantify this time 
lag ($240\ ms$) which is due to the integration time of the optical signal.

The important result at this point is that an IR signal emitted from the APC is 
detected. Noteworthy this signal is detected despite the lower sensitivity of 
optical sensors in the IR range, to be compared with the sensitivity of the 
sensors in the visible range.

\begin{figure}[!h]

\centerline{\includegraphics[width=90mm,height=50mm]{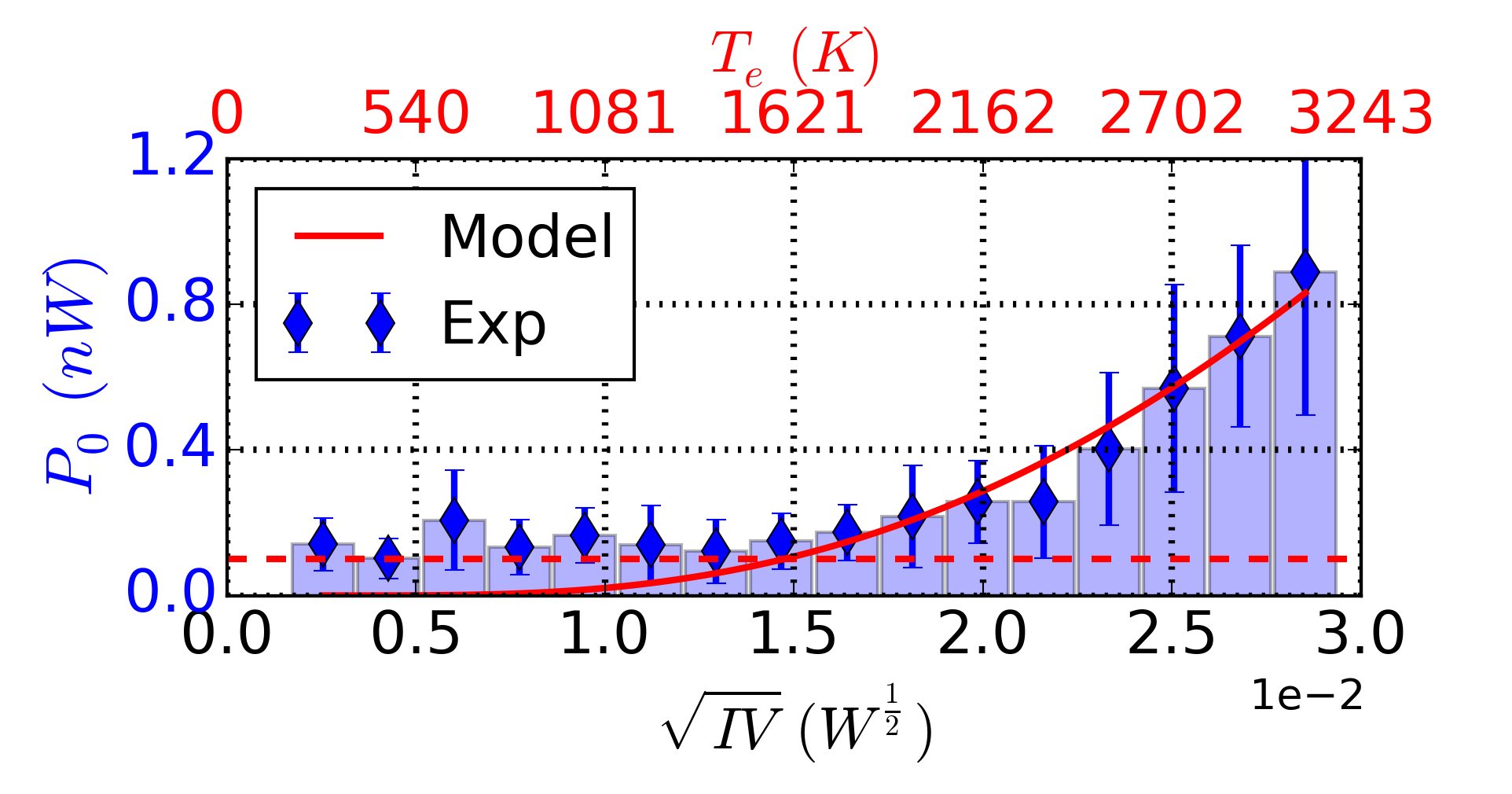}}
\caption{ Experimental (blue columns) and calculated (red continus line) 
dependences of the optical power $P_0$ with $\sqrt{IV}$. Blue columns correspond 
to the mean value of the optical power over a binned x-axis. Error bars show the 
standard deviation calculated over the same binned x-axis. The calculated 
continous line allows the determination of the equivalent hot electrons 
temperature $T_e$ shown on the top x-axis (see discussion section). The dashed 
horizontal red line represents the noise floor of the optical detector.}
\label{fig : IR1}
\end{figure}

Taking into account the time lag, figure \ref{fig : IR1} shows the dependence 
of the optical signal with the electrical power injected in the APC. We 
superimposed (red continuous line), the result of the expected dependence of the 
IR signal modeled assuming black body emission as previously proposed 
\cite{Downes2002}. We will come back to this point in the Discusion section.

The above data is acquired using the full bandwidth of the IR detector. To get 
some basic spectroscopic informations we use semiconductor lowpass filters. The 
smaller the gap the narrower the optical bandwidth. 
Table \ref{tab.1} shows the results acquired in the mW regime ( $V= 0.9\ V$, 
$I=720\ \mu A$).

\begin{table}
\begin{center}
\begin{tabular}{|c|c|}
\hline
Filters & Raw Data (A.U.) \\
\hline
Full Output  & $1.05 \pm\ 0.04$ \\
\hline
Si gap $= 1.12\ eV$ & $0.95\ \pm\ 0.08$ \\
\hline
Ge gap $= 0.68\ eV$ & $0.56\ \pm\ 0.06$ \\
\hline
\end{tabular}
\caption{Optical signal measured using different semi-conductors 
as lowpass filters.}
\label{tab.1}
\end{center}
\end{table}
The data reported in table \ref{tab.1} evidences that half of the signal arises 
from photons with energies lower than the Ge band gap. From this table, by 
difference, we construct table \ref{tab.2} to estimate the proportion of signal 
in each three spectral bands. Table \ref{tab.2} also includes a column of 
computed values that will be described in the discussion.  
\begin{table}[!h]
\begin{center}
\begin{tabular}{|c|c|c|}
\hline
Spectral Bands (eV)  & Measurement & Calculus\\
\hline
Full band $=\intervalleff{\ 0.22\ }{1.2\ }$ & $1$ & $1$ 
\\
\hline
$\intervalleff{\ 0.22\ }{0.68\  }$   & $0.56\ \pm\ 0.06$& $0.61$\\
\hline
$\intervalleff{\ 0.68\ }{1.12\  }$  & $0.37\ \pm\ 0.14$& $0.33$\\
\hline
$\intervalleff{\ 1.12\ }{1.2\  }$ & $0.08\ \pm\ 0.12$& $0.058$ \\
\hline
\end{tabular}

\caption{ Relative optical signals integrated measured and 
calculated over the three spectral bands. Detector sensitivity was taken into 
account.}
\label{tab.2}
\end{center}
\end{table}

Figure \ref{fig : IR2} shows a black-body spectrum ($T=2931\ K$), convoluted by 
the detector spectral response. The three different spectral bands corresponding 
to the use of the optical filters are represented by grey-scale bands below the 
black-body spectrum. Integrating the optical signal for these spectral 
bands allows the calculatation the expected signal reported in the appropriate 
column of table 2.

\section{Discussion}

As mentioned above, APC-LE has been attributed to the radiation from a hot 
electron gas \cite{Downes2002,Schull2009,Buret2015}. The associated observed 
spectrum was proposed to correspond to a black-body like emission from a high 
temperature ($T_{e}$) system\cite{Fedorovich2000,Downes2002,Buret2015}. Such an 
emission spectrum obeys :
\begin{equation}
L(E,T_e) = \frac{2}{(hc)^2}\frac{E^3}{\exp{(\frac{E}{k_bT_e})}-1}
\label{eq : E0}
\end{equation}
with $L(E,T_e)$ being the optical luminance, $T_e$ the temperature, $E$ the 
photon energy, $k_b$ the Boltzmann constant and $c$ the light velocity. Tomchuk 
and Fedorovich showed \cite{Tomchuk1966} that the electronic temperature in 
isolated metal island, with dimension below $L_{e-ph}$, could be related to the 
lattice temperature $T_{L}$ and electrical power following the equation: 

\begin{equation}
(k_BT_e)^2 = (k_BT_L)^2 + \alpha IV
\label{eq : Te}
\end{equation}

Here $I$ is the current and $V$ the applied bias and $\alpha$ an 
empirical constant describing the heating efficiency.

Assuming $ T_L\ <<\ T_e$, we can write :
\begin{equation}
L(E,IV,\alpha) = \frac{2}{(hc)^2}\frac{E^3}{\exp{(\frac{E}{\sqrt{\alpha 
IV}})}-1}
\label{eq : E1}
\end{equation}

Taking into account the spectral response $F(E)$ and bandwidth of the detector, 
integrating the optical luminance and normalizing, we compute the optical power 
$P_0(IV, \alpha)$ :

\begin{equation}
P_0(IV,\alpha) = \int_{E_{min}=0.22\ eV}^{E_{max}=1.2\ eV}{ F(E)L(E, IV ,\alpha) 
} \,\mathrm{d}E
\label{eq : E2}
\end{equation}

The continuous red line of fig.\ref{fig : IR1} is computed from this expression, 
$\alpha$ being the only fitting parameter. The fit was obtained for $\alpha = 
0.014 \hbar$.

From the fitted $\alpha$ and measured $IV$ values we can calculate the 
electronic temperature from $k_BT_e=\sqrt{\alpha IV}$. This $T_e$ values are 
reported on the top axis of fig.\ref{fig : IR1}. In standard operating 
conditions of the MCBJ (i.e. conductance of a few $G_0$), $T_e$ of several 
thousands of degrees, far exceeding $T_L$, are found and fortify the above 
assumptions.

\begin{figure}[!h]
\centerline{\includegraphics[width=90mm,height=50mm]{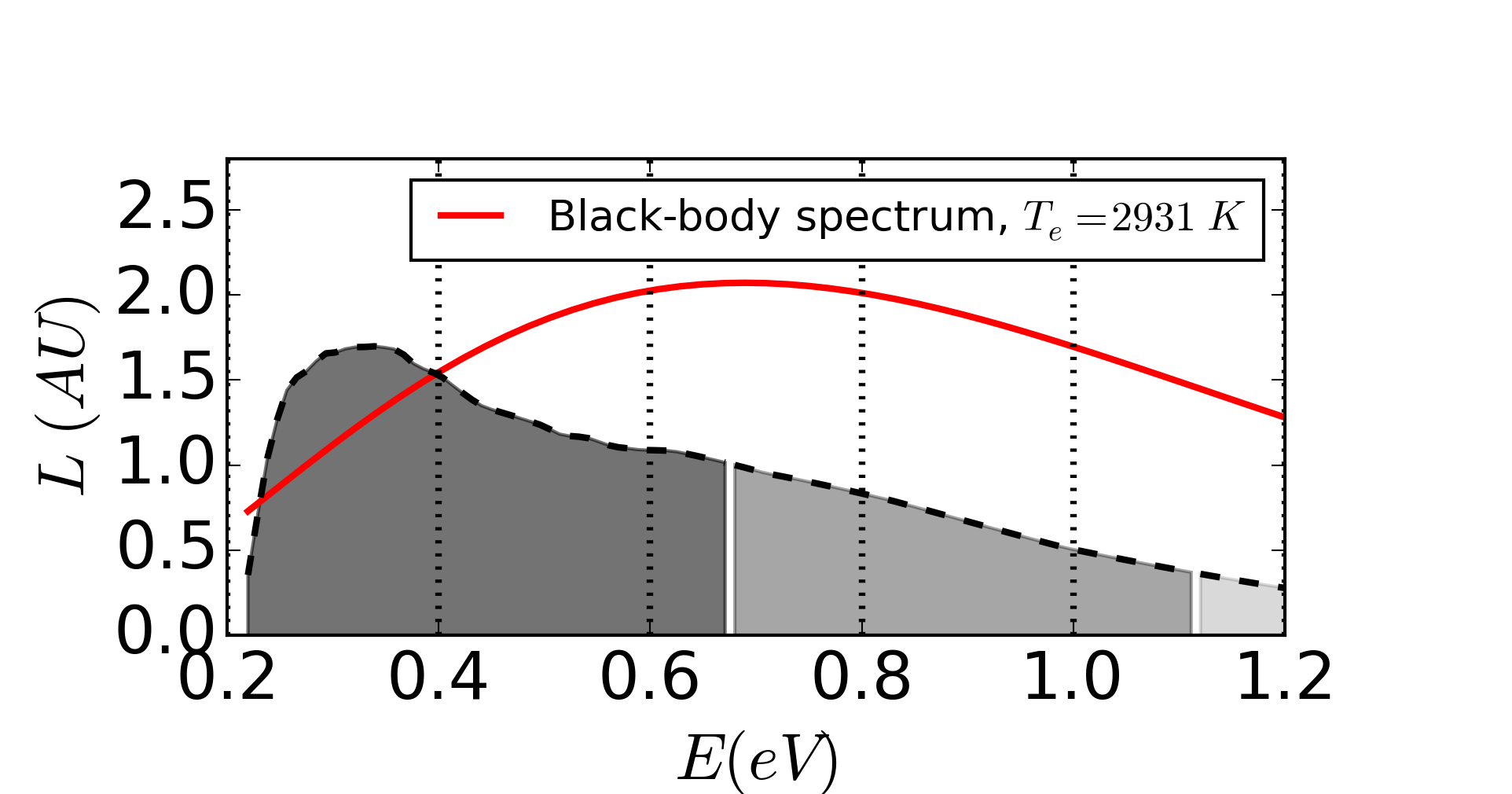}}
\caption{Infra-red spectroscopy using semiconductor as filtre. Red line 
shows the spectrum of a 2931 K Black body. Black dashed line shows this spectrum 
times the detector response. Grey-scale bands indicates the spectral bands used 
to gather integrated optical power.}
\label{fig : IR2}
\end{figure}

From an experiment corresponding to $IV = 1.1\ mW$, knowing the fitted $\alpha $ 
an thus the hot electron gas temperature using eq.\ref{eq : E1} we plot the 
expected black-body spectrum (figure \ref{fig : IR2} dashed curve). 

From it, taking into account the detector response and the properties of the 
three lowpass optical filters, we compute the three expected normalised optical 
powers, using:
\begin{equation}
P(IV,\alpha,E_m,E_M) = \frac{\int_{E_m}^{E_M}{ F(E)L(E,IV,\alpha) } 
\,\mathrm{d}E}{P_0(IV,\alpha)}
\label{eq : E3}
\end{equation}
with $E_m$ and $E_M$ being the lower and upper energy limit of the considerate 
spectral band. These computed values are reported in the last column of 
table.\ref{tab.2}. The agreement with the measured signal is excellent. We point 
out that, although basic, the spectroscopic analysis gives useful results 
corroborating the black-body model of the emission source.\\

Moreover the proportion of signal below and above the germanium bandgap can only 
be consistent with black-body temperatures far above the Au fusion temperature 
of $T_m=1337\ K$.  

\section{Conclusion and Perspectives}

In this letter, we have reported for the first time the observation of IR light 
emission from metallic  point contacts. Results are quantitatively consistent 
with the emission of a hot electron gas which temperature exceeds the melting 
point of gold. They also prolongate the conclusions previously made, at lower 
input power on the basis of the light detected in the visible range, by A.Downes 
et al. \cite{Downes2002} and by M.Buret et al.\cite{Buret2015}. The 
spectroscopic analysis is already very useful and we forecast that in a near 
future the stability of APC at ambient temperature will allow to use more 
advanced and enlightening tools (such as Fourier Transform Infra-Red 
spectroscopy). 

\section*{Acknoledgements}
We thank Mehdi Lagaize for technical support. 


\bibliographystyle{unsrt}
\bibliography{arxiv}

\end{document}